\documentclass[runningheads]{llncs}

\usepackage{graphicx}
\usepackage{amsmath}
\usepackage{listings}
\usepackage{color}

\usepackage{microtype}
\usepackage{wrapfig}

\usepackage{microtype}
\usepackage{booktabs}

\usepackage[]{algorithm2e}

\usepackage{hyperref}

\usepackage{xcolor}
\hypersetup{
    colorlinks,
	linkcolor={blue}, 
	citecolor={blue}, 
	urlcolor={blue}   
}

\usepackage[misc]{ifsym}

\definecolor{codegreen}{rgb}{0,0.6,0}
\definecolor{codegray}{rgb}{0.5,0.5,0.5}
\definecolor{codepurple}{rgb}{0.58,0,0.82}
\definecolor{backcolour}{rgb}{0.95,0.95,0.92}
 
\lstdefinestyle{mystyle}{
    commentstyle=\color{codegreen},
    keywordstyle=\color{magenta},
    numberstyle=\tiny\color{codegray},
    stringstyle=\color{codepurple},
    basicstyle=\footnotesize,
    breakatwhitespace=false,         
    breaklines=true,                 
    captionpos=b,                    
    keepspaces=true,                 
    numbersep=5pt,                  
    showspaces=false,                
    showstringspaces=false,
    showtabs=false,                  
    tabsize=2
}

\newsavebox{\imagebox}

\usepackage{bbding}

\usepackage{cleveref}
\usepackage{subfig}
\usepackage[export]{adjustbox}

\begin{document}
\title{Facilitating Rapid Prototyping in the \\Distributed Data Analytics Platform\\OODIDA via Active-Code Replacement\thanks{The final authenticated version is available online at \url{https://doi.org/10.1016/j.array.2020.100043}.}
}
\titlerunning{Active-Code Replacement in the OODIDA Data Analytics Platform}

\author{Gregor Ulm\inst{1, 2}\Envelope
\orcidID{0000-0001-7848-4883}
\and
Simon Smith \inst{1, 2}
\orcidID{0000-0001-8525-2474}
\and
Adrian Nilsson \inst{1, 2}
\orcidID{0000-0002-8927-845X}
\and
Emil Gustavsson \inst{1, 2}
\orcidID{0000-0002-1290-9989}
\and
Mats Jirstrand \inst{1, 2}
\orcidID{0000-0002-6612-8037}
}
\authorrunning{G. Ulm et al.}

\institute{Fraunhofer-Chalmers Research Centre for Industrial Mathematics,\\ Chalmers Science
Park, 412 88 Gothenburg, Sweden\\
\and Fraunhofer Center for Machine Learning,
\\Chalmers Science
Park, 412 88 Gothenburg, Sweden\\
\email{\{gregor.ulm, simon.smith, adrian.nilsson, emil.gustavsson, mats.jirstrand\}@fcc.chalmers.se}\\
\url{http://www.fcc.chalmers.se/}
}
\maketitle      

\begin{abstract}
OODIDA (On-board/Off-board Distributed Data Analytics) is a platform for distributed real-time analytics, targeting fleets of reference vehicles in the automotive industry. Its users are data analysts. The bulk of the data analytics tasks are performed by clients (on-board), while a central cloud server performs supplementary tasks (off-board). OODIDA can be automatically packaged and deployed, which necessitates restarting parts of the system, or all of it. As this is potentially disruptive, we added the ability to execute user-defined Python modules on clients as well as the server. These modules can be replaced without restarting any part of the system; they can even be replaced between iterations of an ongoing assignment. This feature is referred to as active-code replacement. It facilitates use cases such as iterative A/B testing of machine learning algorithms or modifying experimental algorithms on-the-fly. Various safeguards are in place to ensure that custom code does not have harmful consequences, for instance by limiting the allowed types for return values or prohibiting importing of certain modules of the Python standard library. Consistency of results is achieved by majority vote, which prevents tainted state. Our evaluation shows that active-code replacement can be done in less than a second in an idealized setting whereas a standard deployment takes many orders of magnitude more time. The main contribution of this paper is the description of a relatively straightforward approach to active-code replacement that is very user-friendly. It enables a data analyst to quickly execute custom code on the cloud server as well as on client devices. Sensible safeguards and design decisions ensure that this feature can be used by non-specialists who are not familiar with the implementation of OODIDA in general or this feature in particular. As a consequence of adding the active-code replacement feature, OODIDA is now very well-suited for rapid prototyping.

\end{abstract}

\section{Introduction}
OODIDA~\cite{ulm2019oodida} is a modular system for concurrent distributed data analytics, with a particular focus on the automotive domain. It processes in-vehicle data at its source instead of transferring all data over the network and processing it on a central server. A data analyst interacting with this system uses a Python library that assists in creating and validating assignment specifications which consist of two parts: on-board tasks carried out by the on-board unit (OBU) in a reference vehicle, and an off-board task that is executed on a central cloud server. Several domain-specific algorithms and methods of descriptive statistics have been implemented in OODIDA. However, updating this system is time-consuming and disruptive as it necessitates terminating and redeploying software. Instead, we would like to perform an update without terminating ongoing tasks. We have therefore extended our system with the ability to execute custom code, without having to redeploy any part of the installation. This enables users to define and execute custom computations both on client devices and the server. This is an example of a dynamic code update. With this feature, users of our system are able to carry out their work, which largely consists of either tweaking existing methods for data analytics or developing new ones, with much faster turnaround times, allowing them to reap the benefits of rapid prototyping.

In this paper, we describe the active-code replacement feature of OODIDA. We start off with relevant background information in Sect.~\ref{problem}, which includes a brief overview of our system. In Sect.~\ref{solution} we cover the implementation of this feature, showing how Erlang/OTP and Python interact. We elaborate on the reasoning behind our design considerations, including deliberate limitations, and show how it enables rapid prototyping. Afterwards, we show a quantitative as well as a qualitative evaluation of active-code reloading in Sect.~\ref{eval}, before we continue with related work in Sect.~\ref{related}, plans for future work in Sect.~\ref{future}, and end with the conclusion in Sect.~\ref{conclusion}.

A condensed version of this paper has been previously published~\cite{ulm2019}. That paper presents a quick summary of the active-code replacement feature of the OODIDA platform. In contrast, this paper provides both more depth, covering various implementation details and extensive technical background, as well as increased breadth by giving a more thorough description of relevant parts of our system.

\begin{figure}[]
\centering
\includegraphics[scale=1.0, trim=6.9cm 14.9cm 25.4cm 3.5cm, clip]{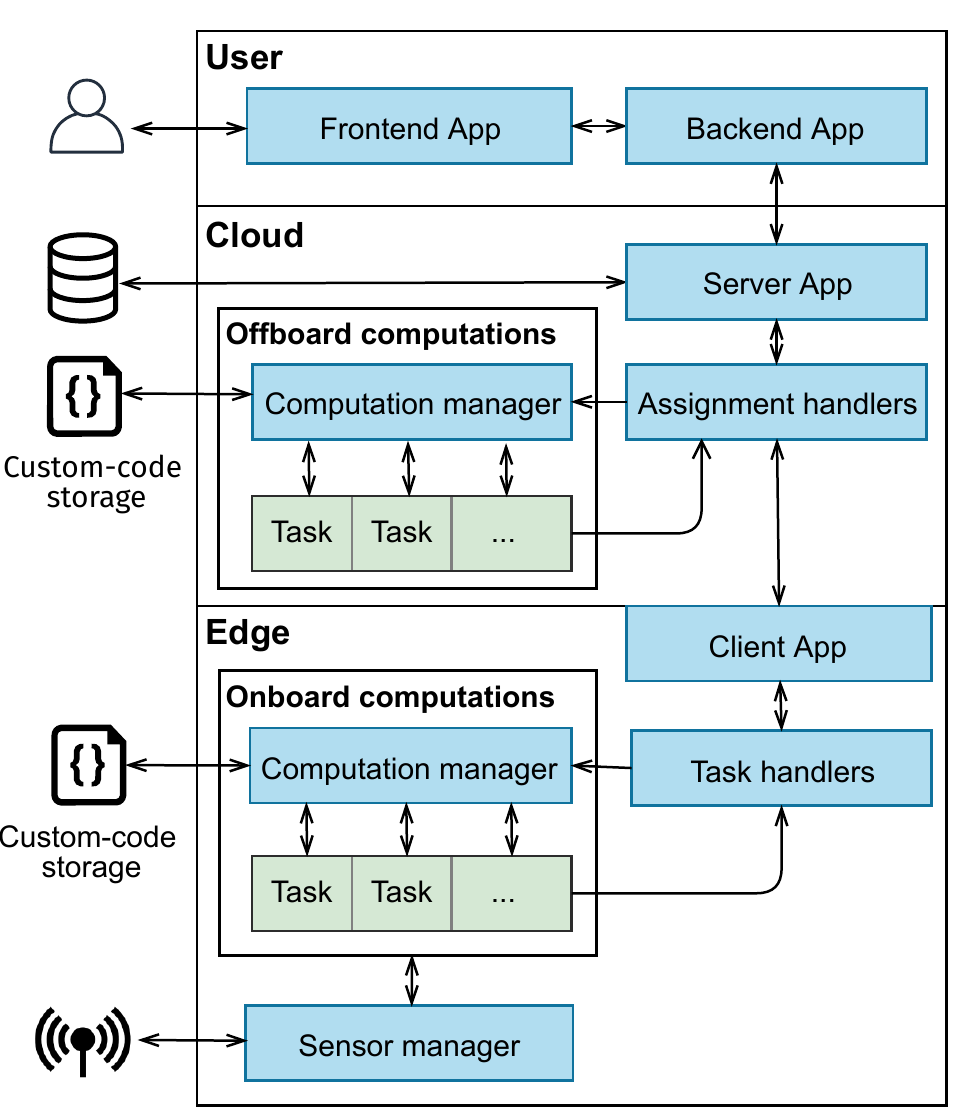}
\caption{OODIDA has a modular architecture. The backend, server, and client apps as well as the assignment and task handlers have been implemented in Erlang/OTP. The user frontend, computation manager and sensor manager have been implemented in Python. Modules or functions that carry out tasks can be implemented in an arbitrary programming language. The sensor manager (bottom) interfaces with the vehicle's CAN bus. Active-code replacement makes use of the custom-code storage depicted on the left.}
\label{fig:architecture}
\end{figure}

\section{Background}
\label{problem}
In this section, we describe the relevant background of active-code replacement in the OODIDA platform. We start with a brief overview of OODIDA~(Sect.~\ref{bg:overview}), including a description of assignment specifications and the user front-end application, before we show how our system can be extended with new computational methods~(Sect.~\ref{bg:extending}). This leads to the motivating use case that describes on-the-fly updating of the system without taking any part of it down~(Sect.~\ref{bg:usecase}).

\subsection{OODIDA Overview}
\label{bg:overview}
This subsection contains a condensed description of OODIDA, which is comprehensively described elsewhere~\cite{ulm2019oodida}. After a brief overview and an example, we highlight some technical details as well as the \emph{status quo ante} of our system for prototyping before the addition of the active-code reloading feature.

\subsubsection{Basic idea.}
OODIDA is a platform for distributed real-time data analytics in the automotive domain, targeting a fleet of reference vehicles. It connects $m$ analysts to $n$ vehicles. The architecture diagram is shown in Fig.~\ref{fig:architecture}. Analysts use OODIDA for data analytics tasks by creating assignments, which are translated into tasks for connected vehicles. These vehicles contain an on-board unit (OBU) that is fit for general-purpose computing. Yet, OBUs are used merely for data analytics. They do not interfere with controlling any part of the vehicle and instead only read CAN bus data. Data analysts use OODIDA for executing various statistical methods and machine learning algorithms. After updating OODIDA in-house, the system can be deployed remotely, which makes new features available to all analysts. A particular focus of this system is on large-scale concurrency: analysts can issue a multitude of tasks to different subsets of clients that are all carried out concurrently. The bottleneck is the available hardware in the vehicles, but experimental results show that we can easily carry out dozens of typical analytics tasks concurrently~\cite{ulm2019oodida}.

The problem our system solves is that data generation of a connected vehicle outpaces increases in bandwidth. There is simply too much data to transfer to a central server for processing. Instead, with OODIDA, data is primarily processed on clients and in real time, which leads to cost savings as transmission and storage costs can be greatly reduced. Furthermore, data analysts can get insights a lot faster, which is valuable for business.

\begin{figure*}[t]
\includegraphics[scale=0.50]{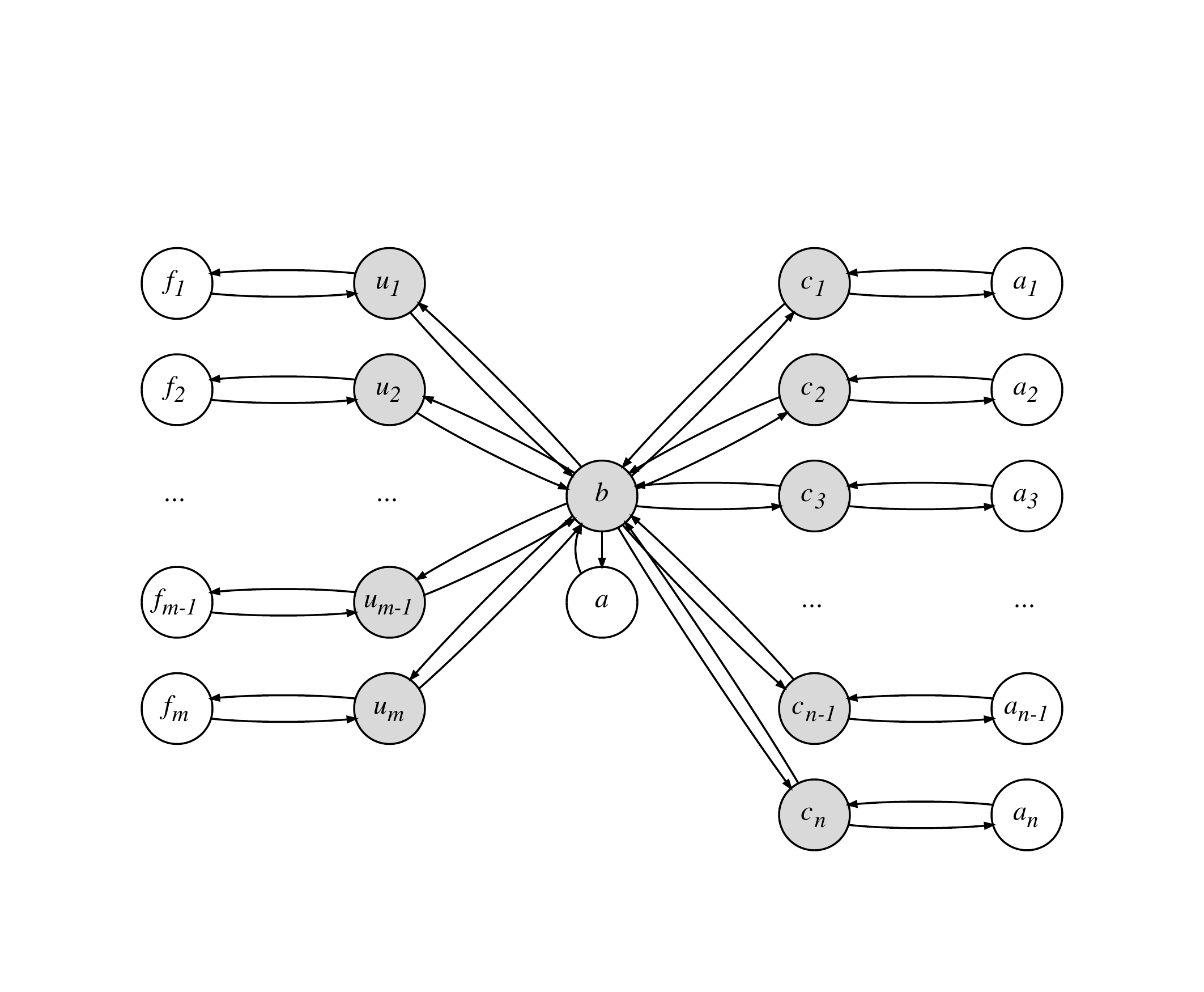}
\caption{OODIDA overview: user nodes $\boldsymbol{u}$ connect to a central cloud $b$, which connects to clients $\boldsymbol{c}$ (vectors in boldface). The shaded nodes are implemented in Erlang/OTP; the other nodes are external Python applications, i.e.\ the user front-ends $\boldsymbol{f}$ and external client applications $\boldsymbol{a}$.}
\label{fig:context}
\end{figure*}

\subsubsection{Technical details}
OODIDA is a distributed system that runs on three kinds of hardware: data analysts use workstations, the server application runs on an internal private cloud, and client applications are executed on OBUs. Our system can accommodate multiple users, but in order to simplify the presentation, we mainly focus on a single-user instance. In Fig.~\ref{fig:context}, the context of our system is shown, indicating that a data analyst uses a front-end $f$. In turn, $f$ is connected to a user module $u$ that communicates with the central cloud application $b$ (bridge). The workstation of the data analyst executes both $f$ and $u$. Node $b$ communicates with client nodes $\boldsymbol{c}$ on OBUs. Each $c$ interacts with an external application $a$. Data analysts use a front-end application $f$ to generate assignment specifications, which are consumed by $u$ and forwarded to $b$. On $b$, assignments are divided into tasks and forwarded to the chosen subset of clients. On client devices, external applications $\boldsymbol{a}$ perform analytics tasks, the results of which are sent to $b$, where optional off-board tasks are performed. Assignments can be executed concurrently.

\begin{figure*}[t]
\includegraphics[scale=0.50]{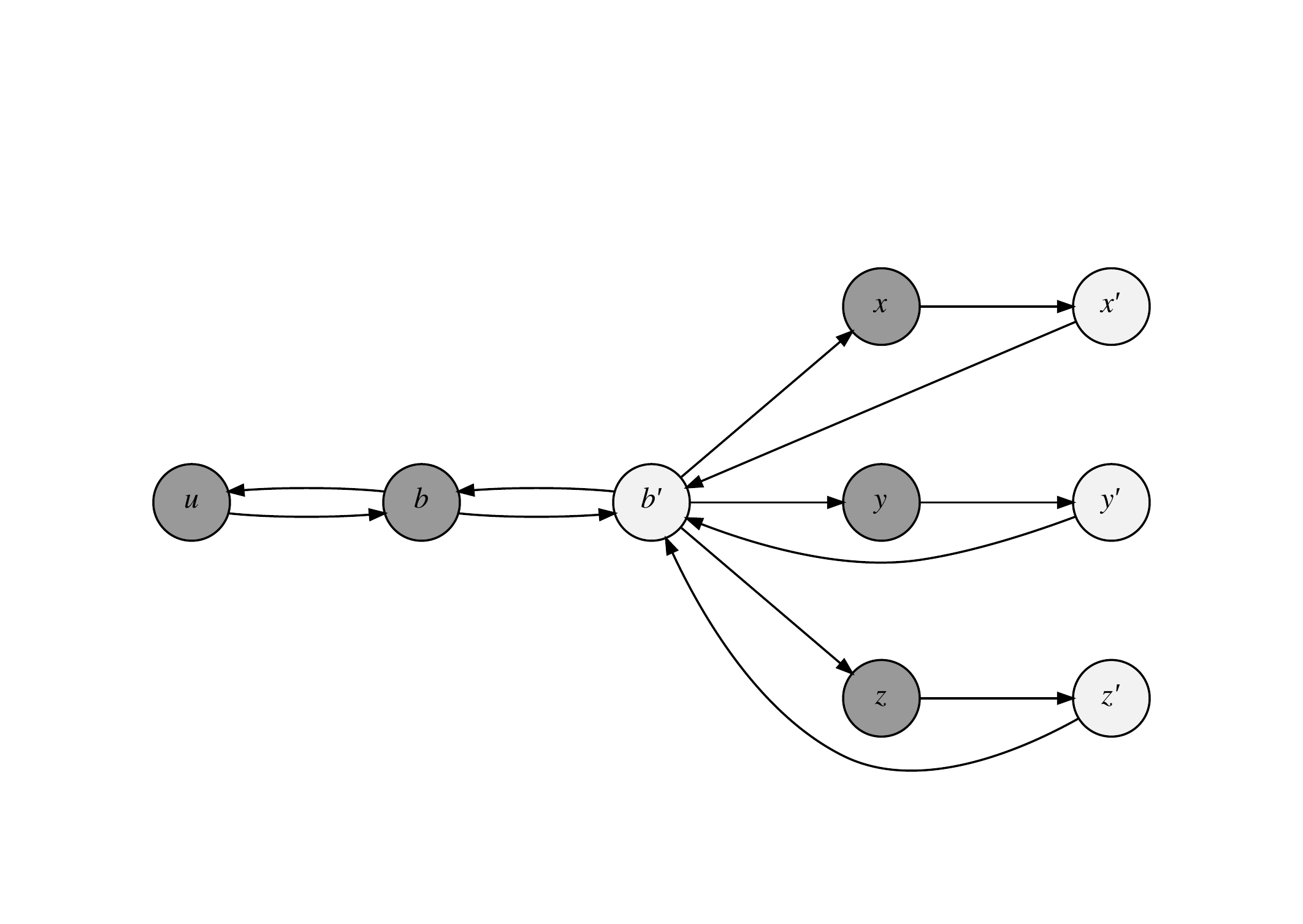}
\caption{This illustration shows the core of OODIDA with permanent nodes (dark) and temporary handlers (light). Clients $x, y$ and $z$ spawn task handlers $x'$, $y'$, and $z'$ that interact with external applications}
\label{fig:whole_fleet_assignment}
\end{figure*}

Building on this more general view, Figs.~\ref{fig:context} and ~\ref{fig:whole_fleet_assignment} present further details of the underlying message-passing infrastructure, which has been implemented in Erlang/OTP. We start with the user node $u$, which is identical to $u$ in Fig.~\ref{fig:context}. The user defines an assignment specification with the help of $f$, which forwards it to $u$. In turn, $u$ forwards it via the network to $b$. That node spawns a temporary assignment handler $b'$, which divides the assignment into tasks and distributes them to client devices. Our illustration shows three client nodes $x, y$ and $z$. Both the client node and its task handler are executed on an OBU, just like the external application $a$ shown in Fig.~\ref{fig:whole_fleet_assignment}. Each client spawns a temporary task handler per received task. For instance, client node $x$ spawns task handler $x'$. Task handlers communicate the task specification to the external application $a$, which performs the requested computational work. Once the results are available, they are picked up by the task handler and forwarded to the originating assignment handler $b'$. Afterwards, the task handler on the client terminates. Once $b'$ has received the results from all involved task handlers, it performs optional off-board computations, sends the results to $b$, and terminates. Finally, $b$ sends the assignment results to the user process $u$, which communicates them to $f$. In which order the various nodes are involved when processing an assignment, including external applications on the client and cloud, is shown in Fig.~\ref{fig:sequence}. The specified on-board and off-board computations can be carried out in an arbitrary programming language as our system uses a language-independent JSON interface. However, we focus on a simplified version of OODIDA that only uses Python applications to execute both on-board and off-board tasks.

\subsubsection{Example Assignment.}
Assignments consist of an on-board task, performed on a central server, and an off-board task, performed by each client that is contained in the selected subset of clients. An example of an assignment specification is provided in Listing~\ref{assignment}, which shows an example of a relatively basic assignment and its definition as an object in Python. The on-board task is executed on the chosen subset of clients, and the off-board task on the central cloud server. The provided example shows an instance of anomaly detection. The entire fleet of vehicles is monitored, with the goal of detecting whenever a vehicle exceeds a speed threshold value of 100. In order to do so, the user specifies the keyword \texttt{collect}, which collects values from the provided list of signals at a certain frequency for a total of $n$ times. In the given example, each client collects 36,000 samples at a frequency of 10 Hz. In total, this means that we monitor each vehicle for a total of 60 minutes. In general, the off-board part of an assignment is relatively inexpensive. While it is possible to perform arbitrary computations on the server as well, it most commonly collects results from clients and forwards them to the user the assignment originated from. This happens in our example as well. However, on top, the keyword \texttt{iterations} with the value '10' is used which indicates that the cloud server will issue the on-board assignment sequentially ten times. Thus, the anomaly detection task will run for ten consecutive hours, with a summary report being sent to the user after each hour.

Lastly, both the on-board and off-board objects are combined into a \texttt{Spec} object for the assignment specification. The user is expected to name such an assignment as well as select a subset of clients, which can be done as random selection of $c$ clients, a numerical selection based on client IDs, selection based on the vehicle model, or, like in our case, as an assignment that is sent out to all clients. In practice the keyword \texttt{all} is only relevant for some assignments because not all signals are available in all vehicles.

\begin{figure}
\centering
\vspace{-1.4em}
\begin{lstlisting}[language=Python, caption=Example of an assignment specification, label=assignment]
onboard = Onboard( 
    computation = 'collect',
    signals     = ['speed'],
    filters     = 'x > 100',
    frequency   = 10  # i.e. 10 Hz
    samples     = 3600 * frequency,  # values of one hour
)

offboard = Offboard(
    computation = 'collect',
    iterations  = 10,  # Run the same assignment 10 times
)

spec = Spec(
    name     = "Sample Assignment",
    clients  = 'all',
    onboard  = onboard,
    offboard = offboard,
)
\end{lstlisting}
\end{figure}

\subsubsection{Prototyping}
 In general, an assignment is a tuple of a chosen algorithm and its parameter values, a subset of signals, and the duration, which is determined by the number of samples and a frequency. This implies that the number of potential assignments is very large but has an upper bound as the number of combinations is finite. Given a reasonably large selection of algorithms to choose from, OODIDA is quite powerful. Yet, a data analyst using this system may want to also deploy novel algorithms. The previously mentioned Listing is a good example: a data analyst interacting with OODIDA may want to start an anomaly detection workflow by filtering out all vehicles that reach a speed of 100 km/h. Yet, this may not be fully sufficient to detect dangerous driving. Consequently, additional criteria may need to be met. For instance, one hypothesis could entail that dangerous driving means driving at a speed of at least 100 km/h for 25\% of the time, but a competing hypothesis could be that driving is only dangerous if it is accompanied by sudden steering angle deviations, or by any steering angle deviation past some threshold. This hints at two needs: first, concurrently testing competing hypotheses and, second, executing computations that may not be definable with the standard methods that are available on our platform.

\subsubsection{User Front-end Application}
OODIDA provides a Python front-end application $f$ for the data analyst for easy creation and validation of assignment specifications. Assignment specifications, an example of which we just detailed, are ultimately turned into JSON objects. Because the manual creation of an assignment as a Python dictionary is error-prone, $f$ automatically verifies the correctness of the provided values. Checks include completeness and correctness of the provided dictionary keys, type-checking the corresponding dictionary values, and verifying that their range is valid. For instance, the value for the field \texttt{frequency} has to be a positive integer and cannot exceed a certain threshold. Separately verifying assignment specifications is necessitated by the rudimentary type system of Python. In programming languages with a more expressive type system, e.g.\ Hindley-Milner type inference or Martin-L{\"o}f dependent types, some of those checks could be performed by the compiler. If the validation is successful, the configuration dictionary is converted into a JSON object and sent to the user process $u$, which is also executed on the workstation of the data analyst.

\subsection{Extending OODIDA}
\label{bg:extending}
As OODIDA has been designed for rapid prototyping, there is the frequent need of extending it with new computational methods, both for on-board and off-board processing. In Fig.~\ref{fig:sequence} a simplified representation of the workflow of OODIDA is given. In short, to extend the system, the worker nodes have to be updated. These are the applications, mostly implemented in Python, that perform on-board and off-board computations. They interact with an arbitrary number of assignment handlers (off-board) and task handlers (on-board). In order to update OODIDA with new computational methods, the system has to be modified. For the user, the only visible change is a new keyword and some associated parameters, if needed. Assuming that we update both the on-board and off-board application, the following steps are required:

\begin{itemize}
\item Update user front-end $f$ to recognize the new off-board and on-board keywords
\item Add checks of necessary assignment parameter values to $f$
\item Add new keyword and associated methods to cloud application worker
\item Add new keyword and associated methods to client application workers
\item Terminate all currently ongoing assignments
\item Shut down OODIDA on the cloud and all clients
\item Redeploy OODIDA
\item Restart OODIDA
\end{itemize}

Unfortunately, this is a potentially disruptive procedure, not even taking into account potentially long-winded software development processes in large organizations. OODIDA has been designed with rapid prototyping in mind, but although it can be very quickly deployed and restarted, the original version cannot be extended while it is up and running. This was possible with \texttt{ffl-erl}~\cite{ulm2019b}, a precursor that was fully implemented in Erlang, which allows so-called hot-code reloading. There are some workarounds to keep OODIDA up-to-date, for instance by automatically redeploying it once a day. As we are targeting a comparatively small fleet of reference vehicles, this is a manageable inconvenience. Yet, users interacting with our system would reap the benefits of a much faster turnaround time if they were able to add computational methods without restarting any of the nodes at all.

\subsection{Motivating Use Cases}
\label{bg:usecase}

\begin{figure*}[t]
\includegraphics[scale=0.50]{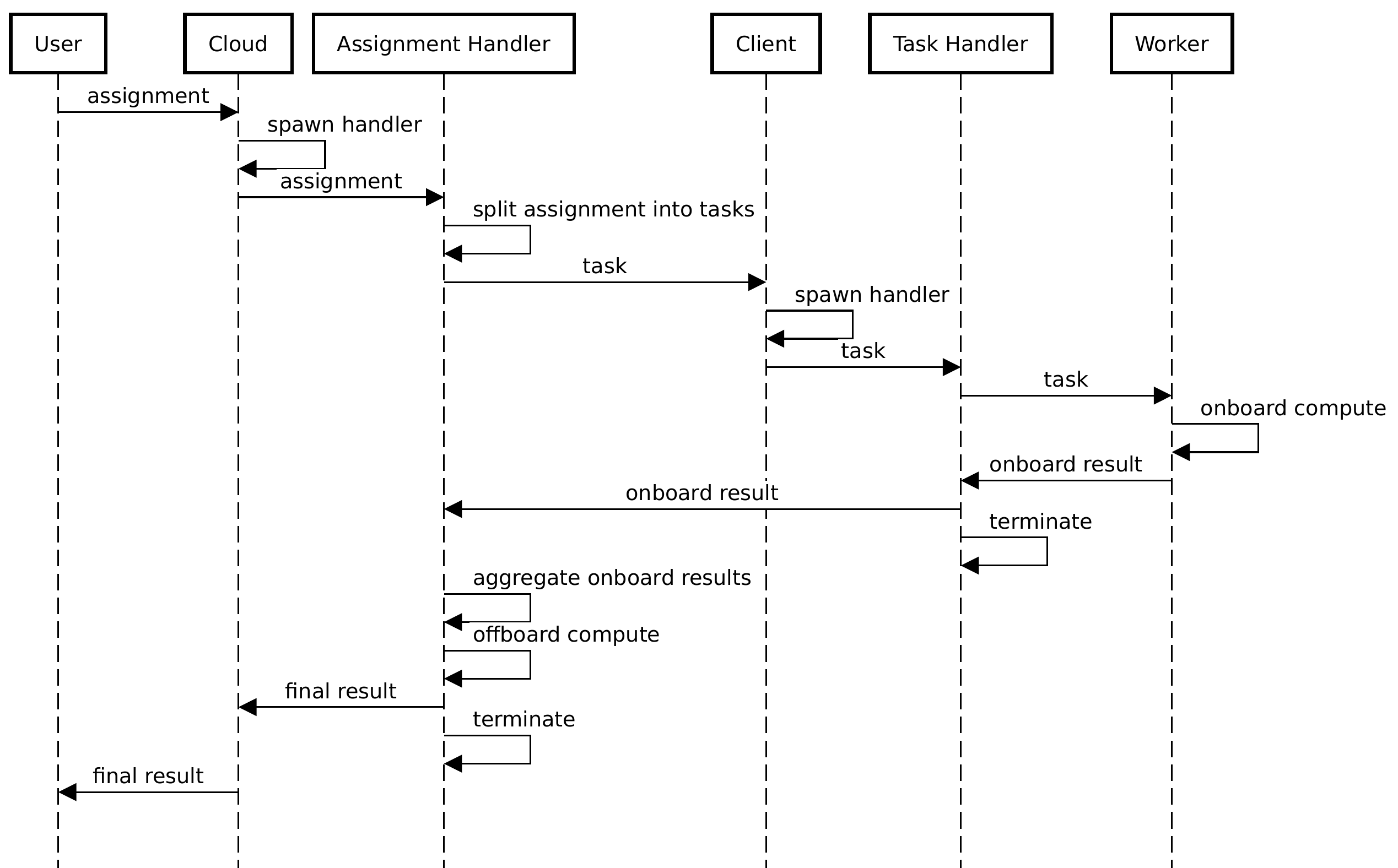}
\caption{Simplified sequence diagram of OODIDA. The system handles multiple users which can all send multiple assignments that can address arbitrary subsets of clients, whereas this diagram only shows one user and one client. On top, clients execute tasks concurrently. The system can also perform iterative work on the task and assignment level. In contrast, this diagram shows an assignment consisting of a task that contains one iteration. The worker node on the cloud is depicted as being part of the assignment handler, which captures the most common off-board use case, i.e.~aggregating results. Refer to Fig.~\ref{fig:context} for a depiction of concurrent workflows.}
\label{fig:sequence}
\end{figure*}

While the previous implementation of OODDIA works very well for issuing standard assignments, there are some limitations. The biggest one is that adding additional algorithms requires updating the worker node on clients or the cloud~(cf.~Fig.~\ref{fig:sequence}). This causes all currently ongoing tasks to be terminated and therefore disincentivizes experimentation. Some tasks may have a runtime of hours, after all. Furthermore, there is the problem that it may not be desirable to permanently add an experimental algorithm to the library on the client. This entails that experimental execution requires two updates, first to push the code update, and afterwards to restore the state before the update.

The first exemplary use case we consider consists of temporarily adding an algorithm to the external client application. If that algorithm proves to be useful, it can be added to all clients via an update of the client software. Otherwise, no particular steps have to be taken as the custom piece of code on the client can be easily deleted or replaced by new custom code. A second, and related use case, consists of temporarily adding different algorithms to non-overlapping subsets of client devices, e.g.\, running two variations of an algorithm, with the goal of evaluating them. Thus, actionable insights can be generated at a much faster pace than the procedure outlined in Sect.~\ref{bg:extending} would allow. 

Lastly, there is the issue of extensibility. Python is a mainstream programming language with a rich ecosystem. There are very comprehensive external libraries available, which are useful for OODIDA, such as the machine learning libraries Keras~\cite{chollet2015keras} and scikit-learn~\cite{pedregosa2011scikit}. As those are vast projects, it is infeasible to create hooks for an entire library. Yet, it is occasionally useful to call a function of those libraries, in which case a data analyst can define a custom-code module that loads the external library and calls that function. Consequently, active-code replacement provides an easy way of quickly accessing the functionality of third-party libraries.

It is also important to keep in mind standard data analytics workflows: Very commonly, analysts write glue code that uses existing libraries. For instance, before running a method with user-selected parameters, data may need to be preprocessed (e.g.~consider the \texttt{sklearn.preprocessing} package). This kind of task is commonly expressed in short scripts. With the feature described in this paper, it is possible to deploy such code to a client. It is very helpful to be able to execute custom preprocessing routines on client devices as this enables new use cases, based on the assumption that it is not feasible to collect data from a large number of client devices in realtime, due to its volume. Another very important task for data analytics workflows is algorithmic exploration that goes beyond merely tuning parameters of algorithms. Instead, this may mean modifying the source code of an existing algorithm or executing algorithms that were written by the analysts themselves. In either case, the code that needs to be deployed tends to be relatively short. As these examples show, it is obvious that the benefit of being able to deploy and execute custom code as opposed to fully redeploying the client software installation leads to a much faster turnaround time. It also enables an entirely different way of working as the ability to quickly deploy custom code heavily encourages experimentation. 

\section{Solution}
\label{solution}
In this section we describe our engineering solution to the problem of replacing active code in our system. We start with the assignment specification the data analyst produces~(Sect.~\ref{sol1}). Afterwards, we focus on the underlying mechanisms for getting a custom piece of code from the data analyst to the cloud as well as client devices~(Sect.~\ref{sol2}). This is followed by discussing implementation details that make it possible to keep devices running while replacing a piece of code~(Sect.~\ref{sol3}), followed by our approach to ensuring consistency of results, based on the fact that not all clients may be updated at the exact same time~(Sect.~\ref{sol4}). Then we highlight security considerations~(Sect.~\ref{sol:security}). Afterwards we show how a complex use case can be implemented with active-code replacement~(Sect.~\ref{sol5}) and discuss deliberate limitations of our solution~(Sect.~\ref{sol6}).

\subsection{Using Custom Code in an Assignment}
\label{sol1}
In line with the guiding principle that OODIDA should make it as easy as possible for the data analyst to do their job, active-code replacement has been designed to minimize the need for interventions. The data analyst only has to carry out two steps. The first is providing a stand-alone Python module with the custom code. It could include imports, which, of course, the user has to ensure to be available on the target OBUs. The only requirement for the structure of the custom code module is that it contains a function \texttt{custom\_code} as an entry point, which takes exactly one argument. This is the function that is called on the cloud or client. Additional parameters have to be hard-coded. Before being able to call custom code, it needs to be deployed. To do so, the user needs to specify the location of the code file on their machine and afterwards call the function \texttt{deploy\_code}, which takes as an argument the target, \texttt{onboard} or \texttt{offboard}, the location of the file and optionally a specification of the intended clients, which is ultimately a list of client IDs. In Listing~\ref{lst:custom} we call a helper function to retrieve the IDs of all vehicles of a particular type. It is possible to send different modules to non-overlapping subsets of clients via subsequent assignments.

The verification process of the user front-end application consists of two steps. First, the provided module has to be syntactically correct, which is done by loading it in Python. The second check targets the prescribed function \texttt{custom\_code}. That function is called with the expected input format, depending on whether it is called on the client or the cloud. We also verify that the returned values are of the expected type. If any of these assertions fail, the assignment is discarded. Otherwise, the custom Python module is sent to the cloud or to clients, depending on the provided instructions. This step is preceded by producing another assignment specification that, in either case, contains the entries \texttt{user\_id} and \texttt{custom\_code}. The value of the latter is an encoding of the user-provided Python module. The value of the key \texttt{mode} is either \texttt{deploy\_offboard} or \texttt{deploy\_onboard}. Once custom code has been deployed, it can be referred to in assignments by setting the value of the keys \texttt{onboard} or \texttt{offboard} to \texttt{custom}.

\begin{lstlisting}[language=Python, caption={Deploying custom on-board code}, label={lst:custom}]
codefile = 'framework/custom_functions/histogram_onboard.py'
clients = = user.get_models('type_a')

user.deploy_code('onboard', codefile, clients)
onboard = Onboard(
  computation = 'custom',
  signal = 'speed',
  samples = 120,
  frequency = 5,
)

hist_spec = Spec(
  name     = "Histogram example",
  clients  = clients,
  onboard  = onboard,
  offboard = Offboard('collect'),
)
\end{lstlisting}

\subsection{Code Forwarding}
\label{sol2}
Assuming the provided custom code for the client has passed the verification stage, it is turned into a JSON object and ingested by the user module for further processing. Within that JSON object, the user-defined code is represented as an encoded text string. The user module extracts all relevant values from the provided JSON object and forwards it to the cloud process $b$. In turn, $b$ spawns a new assignment handler $b'$ for this particular assignment. The next step depends on whether custom code for the server or client devices has been provided. 

The process of turning an assignment into tasks for client devices does not depend on the provided values and is thus unchanged from the description in the paper on OODIDA~\cite{ulm2019oodida} or the brief summary presented earlier in this paper. Node $b'$ breaks the assignment specification down into tasks for all clients specified in the assignment. After this is done, task specifications are sent to the designated client processes. Each client process spawns a task handler for the current task. Its purpose is to monitor task completion, besides alleviating the edge process from that burden and enabling it to process further task specifications concurrently. In our case, the task handler sends the task specification in JSON to an external Python application, which turns the given code into a file, thus recreating the Python module the data analyst initially provided. The name of the resulting file also contains the ID of the user who provided it. After the task handler is done, it notifies the assignment handler and terminates. Similarly, once the assignment handler has received responses from all task handlers, it sends a status message to the cloud node and terminates. The cloud node sends a status message to inform the user that the custom code has been successfully deployed. Deploying custom code to the cloud is similar, the main difference being that $b'$ communicates with the external Python worker application running on the cloud.

\subsection{Code Replacement}
\label{sol3}
Computations are performed only after the specified amount of data has been gathered. This implies that a custom code module can be safely replaced as long as data collection is ongoing. The case where an update collides with a function call to custom code is discussed in Sect.~\ref{sol4}. If a custom on-board or off-board computation is triggered by the keyword \texttt{custom}, Python loads the user-provided module using the function \texttt{reload} from the standard library. This happens in a separate process using the \texttt{multiprocessing} library. The motivation behind this choice is to enable concurrency in the client application as well as to avoid some technical issues with reloading in Python, which would retain definitions from a previously used custom module. Instead, our approach creates a blank slate for each reload.

The user-specified module is located at a predefined path, which is known to the reload function. Once loaded, the custom function is applied to the available data in the final aggregation step, which is performed once and at the end of a task or assignment. When an assignment using a module with custom code is active, the external applications reload the custom module with each iteration. This may be unexpected, but it leads to greater flexibility. Consider an assignment that runs for an indefinite number of iterations. As the external applications can process tasks concurrently, and code replacement is just another task, the data analyst can, for instance, react to intermediate results by deploying custom code, with modified algorithmic parameters, that is used in an ongoing assignment as soon as it becomes available. As custom code is tied to a unique user ID, there is furthermore no interference due to custom code deployed by other users as every unique user ID is tied to a unique user account, and each user of the system has their own account.

One theoretical issue with our approach is that modules may be reloaded repeatedly, which is inefficient. Yet, OODIDA was not designed with the idea of running arbitrary libraries on the client; instead, it has a strict focus on distributed data analytics. This entails that external libraries do not pose a problem as bread-and-butter libraries such as \texttt{scikit-learn} and Keras are loaded already when the client application is started. Thus, these modules are available to a custom module and should not be imported again by custom code; such imports are reported by the user-side validator. On top, the user does not have the ability to deploy additional libraries by themselves. Instead, they can only access the Python standard library, with some limitations, and a small set of third-party libraries. Consequently, code that is imported with each iteration tends to be small and does not depend on additional external libraries.

\subsection{Ensuring Consistency}
\label{sol4}
Inconsistent updates are a problem in practice, i.e.~results sent from clients may have been produced with different custom code modules in the same iteration of an assignment. This happens if not all clients receive the updated custom code before the end of the current iteration. In a streaming context, where clients have the ability to peek into results to get intermediate updates, the same issue could emerge, namely that clients use different versions of custom code for their computations. To solve this problem, each provided module with custom code is tagged with its md5 hash signature. This signature is reported together with the results from the clients. The cloud only uses the results tagged with the signature that achieves a majority and discards all others. Consequently, results are never tainted by using different versions of custom code in the same iteration. The expectation is that any updated custom code would eventually, and quickly, reach a majority. An update may not succeed for various reasons. If it is because a client has become unavailable, then said client cannot send any results anyway. Should an update not succeed, then the client reports an error. In that case, the update has to be repeated.

It is possible that a new custom code version arrives at the same time the client application wants to load it. This is one example where the standard approach would be to roll back the update and instead use the previous custom code version. However, this scenario is less of a concern for us as we replace computational methods instead of system-level software. There is deliberately no mechanism for a rollback as the old version of the custom code was supposed to be replaced by new custom code, which implies that any results that could be generated by the old code instead of the not-yet-available new one are not of any interest to the user anymore. The only exception where a rollback would be helpful is in the pathological case where the old and new version of the provided custom code are identical, and the update with the new custom code failed. Consequently, we deliberately let this one computation fail and the client report an error. For the next iteration, the new version of the custom code can be expected to be available, which means that this issue resolves itself quickly in practice.

\subsection{Security Measures}
\label{sol:security}
The design of our system addresses both internal and external threats, both accidental and deliberate ones. We consequently limit the expressibility of the code the user can deploy. In addition, OODIDA is designed to run on a corporate VPN.

First, in order to limit the damage the user can do, we enforce that the provided custom function takes a list of numerical values as input and returns either a list of numerical values or a numerical value. Verifying this property is carried out by the user-front end application before dispatching a piece of custom code for further processing. There are corresponding assertions on the on-board and off-board nodes as it is theoretically possible to manually specify custom code that sidesteps the checks in the front-end application. However, this would require knowledge of the implementation of those checks, some of which use randomly generated inputs and others static ones. Thus, we consider it highly unlikely that an antagonistic developer would be able to work around the checks in the user front-end. The user would have to reliably predict the input of the test cases and cover it with branching logic, which is triggered in that particular case, but not otherwise. Guessing randomized inputs is arguably not feasible. However, even with perfect knowledge of the implementation and the seed used for generating random data, an antagonistic user would be thwarted. The reason is that we programmatically ensure proper behavior of the custom function that is executed on the client via a \texttt{try-except} construct for error handling. This step also includes a verification that the returned values are indeed as specified. This would seem like duplicate work as this check is already performed in the front-end prior to deployment, but it addresses the case of an omniscient antagonistic user. Thus, this approach closes the previously mentioned loophole an antagonistic developer theoretically has on the client side. A caveat is that this relies on the provided function terminating. Yet, there is the issue of the halting problem, i.e.\ an antagonistic developer could write a function that never returns, thus wasting computational resources of the client. Some legitimate computations could take a significant amount of time, so it is not possible to distinguish between legitimate and antagonistic code on that metric alone. However, with a generous timeout clause, which is enabled via Python's \texttt{multiprocessing} library, a custom function that exceeds a set threshold, regardless of whether the code is antagonistic or not, can easily be terminated. Therefore, we tackle the problem of antagonistic input sufficiently well from a practical perspective. Also, it has to be kept in mind that our system supports commercial users who can be assumed to be invested in the system being fully operational.

Second, there is the issue of cloud security. As has been pointed out, commercial cloud solutions are vulnerable~\cite{shaikh2011security,islam2016classification} and therefore require adequate responses~\cite{sabahi2011cloud}. Among others, there are vulnerabilities due to multitenancy, virtualization, and resource-sharing. As we were aware of those issues, we chose to sidestep them by executing OODIDA on a corporate VPN instead. This does not mean that network security is not an issue. However, this approach avoids additional threats that are unique to cloud computing. On a related note, the primary motivation behind this choice was the need to protect our data. In that regard, not relying on a third-party commercial cloud computing provider seemed like an obvious decision.

\subsection{Complex Use Cases}
\label{sol5}
The description of active-code replacement so far indicates that the user can execute arbitrary code on the server and clients, as long as the correct inputs and outputs are consumed and produced. What may not be immediately obvious, however, is that we can now even create \emph{ad hoc} implementations of the most complex OODIDA use cases, an example of which is federated learning~\cite{mcmahan2016communication}. A key aspect of federated learning, compared to many standard types of assignments on our system, is that the results of one iteration are used as the input of the next one. The original implementation is discussed in the paper on OODIDA~\cite{ulm2019oodida}. With federated learning, clients update machine learning models, which the server uses as inputs in order to create a new global model. This global model is the starting point for the next iteration of training on clients.

\begin{figure}
\centering
\vspace{-1.4em}
\begin{lstlisting}[language=Python, caption=Example of an assignment specification, label=lst:fl]
onboard = Onboard(
    computation = 'custom',
    parameters  = {'result_flow': 'connected'}  # Result of offboard goes back to the clients
)

offboard = Offboard(
    computation = 'custom',
    iterations  = 3
)

fl_spec = Spec(
    name     = "Custom Federated Learning",
    clients  = 3,
    onboard  = onboard,
    offboard = offboard,
)\end{lstlisting}
\end{figure}

Assignments using in-built functions have parameters attached to them that the user can tweak. For custom code, we added a parameter for controlling the workflow called \texttt{result\_flow}, which can take two values: \texttt{isolated}, which is the default, and \texttt{connected}. With the latter, the results of one iteration are used as input for the next iteration. Listing~\ref{lst:fl} shows the exemplary use of it, where code running on the client retrieves a model. Custom code containing an implementation of an artificial neural network was deployed to clients. In addition, custom code for averaging the received updated local models was deployed to the server. Clients produce these local models at the end of each iteration and send them to the server. The available parameters for custom code are user-definable, except that the keyword \verb|result_flow| is reserved. For custom code that uses a \verb|connected| workflow, the initial value for the model is set to \verb|null|. The \verb|parameters| argument can be used in a flexible manner. For instance, in an assignment built on using federated learning, an initial global model might be desirable, which could be included in this argument.

\subsection{Limitations}
\label{sol6}
We consider custom modules a temporary solution. Any function that is deemed generally useful should be added permanently to worker nodes in an update. We also want to discourage data analysts from heavily relying on custom functions. Currently, they can each define one custom-code module for the cloud, and one for each client. It would be straightforward to add the ability to handle multiple custom-code modules per user. Yet, we do not want multiple custom functions to be part of their regular workflow as the consequences of this would be undesirable. In the worst case, this would mean that each user of the system maintains their own custom functions instead of building up a library with relevant algorithmic methods for all users. There is also a practical aspect of why we want to discourage users from continually using custom functions instead of permanently adding them to the client installation in the form of a library. The client application is designed to be stateless: it processes sensor data, discards these data after processing, and sends results to the server. With this approach, it is straightforward to execute experiments and also replicate them. In contrast, replication with custom code would require to separately keep track of the source code files that were deployed, and in order to do so effectively and reliably, one would need to build an additional component for OODIDA that automates this, in order to reduce human error.

As previously stated, users only have the ability to deploy one source code file. For security reasons, we chose to not implement the ability to deploy additional libraries \emph{ad hoc}. The standard data science libraries we use are very well established, which is why they are installed on each client. For standard tasks, which consist of the application of existing methods, this is not a limitation. However, users may need to develop new algorithmic methods. While it is not possible to upload a modified \texttt{scikit-learn} library module, there is a very effective workaround for this situation: It is possible to include library source code in the custom-code module the user wants to deploy. Of course, some modifications may be needed to make the user-defined functions callable within the very same file. In any case, this is an adequate workaround when the goal is to, for instance, test a new algorithm in practice. Granted, normally such an algorithm would be part of an external library, but as long as it is being worked on, it is simply included in the deployed file. Lastly, we would like to add that while users cannot add their own libraries to the client, administrators can. Clients can even be remotely updated, so the limitations users face is a deliberate design decision instead of an unintended limitation of our system.

\section{Evaluation}
\label{eval}
In this section we present a qualitative and quantitative assessment of active-code replacement. The goal is to both quantify the benefits of our approach as well as argue for its soundness. We start with the design of the experiment~(Sect.~\ref{eval:design}), which is followed by a description of the data we used (Sect.~\ref{eval:data}) and a note on the environment that was used for the evaluation~(Sect.~\ref{eval:setup}). Afterwards, we show our results and discuss them~(Sect.~\ref{eval:results}).

\subsection{Experimental Design}
\label{eval:design}
The main benefit of active-code replacement is that code for new computational methods can be deployed right away and executed almost instantly, without affecting other ongoing tasks. In contrast, a standard update of the cloud or client installation necessitates redeploying and restarting the respective components of the system. In order to quantitatively evaluate the performance difference, we executed OODDIA in an idealized scenario where the user and server were executed on one workstation each and one client device on another. We deployed identical code to both the client and the server and took the average of five runs. The standard deployment procedure assumes that a minimal Linux installation with all necessary third-party applications and libraries is available. A complete re-installation would take considerably more time. The total amount of data that needs to be transferred in order to deploy OODIDA is around 1 MB. This is zipped data.

\subsection{Data Description}
\label{eval:data}
Custom code that is to be deployed remotely is turned into a payload for an assignment, which is a lightweight JSON format. We took a standard real-world example consisting of 20 lines of Python source code. The source code file has a size of 0.45 KB. When turned into the payload of a JSON assignment specification, the size of the file to be sent via the network is around 0.52 KB. The amount of custom code used may seem like a rather small amount of code, but this is beyond typical assignments that consist largely of glue code, tying together various calls to \texttt{scikit-learn} methods. These are often in the order of 10 lines of code or even below that.

\subsection{Hardware and Software Setup}
\label{eval:setup}
 We used three quad-core PC workstations. Workstation 1 contains an Intel Core i7-6700k CPU (4.0GHz), workstation 2 and 3 an Intel Core i7-7700k CPU (4.2GHz). These workstations are equipped with 32 GB RAM each, of which 23.5 GB were made available to the hosted Linux operating system. They run on Windows 10 Pro (build 1803) and execute OODIDA on Ubuntu Linux~16.04 LTS in VirtualBox 6.0. Workstation 1 executed the one instance of the user front-end, workstation 2 the cloud application of our system, and workstation 3 one instance of a client application. These workstations were connected via Ethernet.

\subsection{Results and Discussion}
\label{eval:results}
In our idealized test setup, where the various workstations that run the user, cloud and client components of OODIDA are connected via Ethernet, it takes a fraction of a second for a custom on-board or off-board method to be available for the user to call when deployed with active-code replacement, as shown in Table~\ref{tab:comparison}. On the other hand, automated redeployment of the cloud and client installation takes roughly 20 and 40 seconds, respectively. The runtime difference between a standard update and active-code replacement amounts to three orders of magnitude. The results of the standard deployment do not take into account organizational processes that may artificially lengthen deployment times. Thus, a standard deployment does, in the real world, not just take around 20 and 40 seconds, respectively, but those times plus an additional $\Delta$, which is incomparably larger.

\begin{table}[]
\centering{
\caption{Runtime comparison of active-code replacement of a moderately long Python module versus regular redeployment in an idealized setting. The provided figures are the averages of five runs. As the numbers show, the former has a significant advantage. Yet, this does not factor in that a standard update is more invasive but can also be more comprehensive. This is expressed by the additional quantity $\Delta$, which captures delays due to industry best-practices but also bureaucracy in organizations. A standard deployment can easily take days or weeks, depending on the used software development practices.
 \\}
\label{tab:comparison}
\begin{tabular}{@{}lll@{}}
\toprule
& Cloud & Client \\ \toprule
Active-code replacement \phantom{aaaaaaa}
&  20.3 ms   \phantom{aaaaaaa} & 45.4 ms        \\
Standard redeployment
& 23.6 s + $\Delta$ &  40.8 s + $\Delta$\\
\bottomrule
\end{tabular}
}
\end{table}

Our comparison also highlights that active-code replacement is less bureaucratic and less intrusive as it does not require interrupting any currently ongoing assignments. Indeed, in a realistic industry scenario, an update could take days or even weeks due to software development and organizational processes. The big benefit of active-code replacement is that the user, a data analyst, can issue updates themselves. In contrast, a standard deployment depends on software engineers committing code. Even in a continuous deployment environment in a fast-moving agile software development process, the organizational overhead between defining the specification of an additional method that is to be added to the cloud or client installation of OODIDA and the final merging can take days, taking into account code-review and best practices like merging new commits only after the approval of other team members. Consequently, the difference $\Delta$ between the milliseconds it takes to carry out active-code replacement and a standard deployment adds many orders of magnitude more to the measured time difference. 

A justification for our idealized test setup is in order. Of course, real-world deployment via a wireless or 4G connection would be slower as well as error-prone. Yet, our evaluation environment reveals the relative performance difference of both approaches, eliminating potentially unreliable data transmission as a source of error. In fact, given that the standard deployment procedure requires the transmission of a much larger amount of data via the network, it would be disadvantaged in an experiment via wireless or 4G as the potential for packet losses increases due to the greater amount of data used. Thus, our experimental results underestimate the true difference in deployment speed. That being said, the largest factor will be due to $\Delta$, whereas the actual time for deployment is negligible in contrast.

Despite the benefits of active-code replacement we just mentioned, it is not the case that this approach fully sidesteps the need to update the library of computational methods on the cloud or on clients as OODIDA enforces restrictions on custom code. For instance, some parts of the Python standard library are off limits. Also, the user cannot install external libraries. Yet, for typical algorithmic explorations, which users of our system regularly conduct, active-code replacement is a vital feature that increases user productivity far more than the previous comparison may imply. That being said, due to the limitations of active-code replacement, it is complementary to the standard update procedure rather than a competitive approach.

\section{Related Work}
\label{related}
The engineering challenge described in this paper is an extension of the OODIDA platform~\cite{ulm2019oodida}. It was originally an Erlang-only system based on \texttt{ffl-erl}, a framework for federated learning~\cite{ulm2019b}.

Active-code replacement is a niche topic in applied computing, despite its use in fault-tolerant systems. That it is a niche topic is obvious from the fact that there is no uniform terminology. For instance, the official Erlang documentation calls it \emph{code replacement}.\footnote{Refer to the section "Compilation and Code Loading" in the official Erlang documentation: \url{http://erlang.org/doc/reference_manual/code_loading.html} (Accessed March 4, 2019)} Among working Erlang programmers, the terms \emph{hot-code loading} or \emph{hot-code swapping} are more common. In contrast, the standard textbooks by Cesarini et. al~\cite{cesarini2009erlang,cesarini2016designing} refer to it as \emph{upgrading processes}. In academia, we encounter further names for this concept. In an extensive survey by Seifzadeh et al.~\cite{seifzadeh2013survey} the chosen term is \emph{dynamic software updating}. According to their terminology, our approach is called \emph{updating references of all dependants}. Giuffrida and Tanenbaum~\cite{giuffrida2009prepare,giuffrida2010taxonomy} call it \emph{live update}, further classified as \emph{changes to code}. Of course, given that the terminology is not settled, we saw little reason for not creating a term we considered most suitable for describing the actions that take place.

In terms of descriptions of systems that perform active-code replacement, several approaches have been suggested in the literature. We focus on a selection of them and provide a qualitative comparison with active-code replacement in OODIDA. Concretely, we chose MUC by Qiang et al.~\cite{QIANG2017}, Polus by Chen et al.~\cite{chen2007polus}, and Javelus by Gu et al.~\cite{gu2012javelus}. Abstractly viewed, these approaches share some similarities with active-code replacement but they solve different problems. The most significant difference is that our approach enables temporary updates of a specific part of the application, which is due to our use case of distributed data analytics, as described earlier. We did not find a mechanism for distributed active-code replacement that is quite comparable to ours as we focus on letting the user execute custom code in a restricted environment. In contrast, the various update mechanism we found in the literature focus on updating parts of the application that is being executed, often seemingly without any restrictions. It would be far too unsafe to do that in an \emph{ad hoc} manner, which underscores the difference between our approach and others.

Starting with MUC, the major difference is that this system exclusively targets cloud-only updates, while active-code replacement also addresses a potentially very large number of client devices. In MUC, each update leads to forking a new process that is executed concurrently. These processes are synchronized to ensure consistency. In contrast, in OODIDA, assignments are executed concurrently. Yet, as there is no continual processing, our problem is simplified. We thus replace code at any point but only execute it the next time it is called, which is after a batch of results has arrived. There is no need to concurrently execute an old version of the code as it would taint the state a computation if we combined results that were arrived at with non-identical code.

A significant difference between active-code replacement in OODIDA and Polus is that the latter replaces larger units of code instead of isolated modules. As mentioned above, this is a much more wide-ranging procedure than updating custom code for data analytics. In addition, Polus operates in a multi-threading environment instead of the highly concurrent message-passing environment of OODIDA. This is a fundamental difference, which means that the approach of Polus could not be easily replicated in OODIDA. In essence, we spawn a separate process that uses custom code. In order to replicate the approach of Polus, we would need to update the existing process, which would make sandboxing much more difficult. In our case, a spawned process handler that crashes due to illegal custom code, which should not happen to begin with, cannot take down the entire system. Instead, this process is terminated and the custom code discarded.

The goal of Javelus is updating a stand-alone Java application as opposed to a distributed system. This is only indirectly related to our use case. However, in principle, one could of course assume that this application is executed on the cloud and assume that mechanisms that enable remote updates are in place. We only mention Javelus because it uses an approach that is quite similar to ours. Via a "lazy update mechanism", code replacement only has an effect if a module is indeed used. While there are arguably some issues with this approach in a stand-alone application where such updates are intended to not only be temporary, this is elegant nonetheless. In active-code replacement, we do something very similar. By separating custom-code deployment and custom-code execution, we only load a custom module if it is needed. In fact, this solution was suggested by the fact that our system first gathers data and afterwards processes it. Even streams are emulated as micro-batches. The observation that code that performs data analytics tasks is only invoked in fixed intervals subsequently led to the creation of the feature described in this paper. The key modification we had to make was to make it possible to swap such a piece of code, either by pointing to different code or by updating a file that uses the same reference. We chose the latter approach as it was more straightforward to implement, even though there would be little difference in practice, had we chosen the other approach.

\section{Future Work}
\label{future}
In addition to plans for future work on the OODIDA platform in general, we also have concrete ambitions for active-code replacement. The feature, as described, works as intended. Yet, we did point out some limitations and workarounds. The guiding idea is that the execution of custom code should be safe. One way of achieving this is by placing restrictions on the provided code, such as only allowing certain return types. As of now, we only allow numerical values or lists of numerical values. This is sufficient for a very large number of data analytics tasks. Yet, in order to make active-code replacement more useful, we should extend the set of allowed return types. Related is the problem that we exclude certain Python libraries from being called. A further investigation is needed in order to determine if a more fine-grained approach would make sense.

A major limitation of the current iteration of active-code replacement in OODIDA is that the user is not allowed to deploy additional libraries. However, as this is a potentially very useful feature, we would like to explore the possibility of creating sandboxed environments on the client, perhaps a solution based on lightweight containers via Docker~\cite{merkel2014docker}. This also refers back to an earlier note (cf.~Sect.~\ref{sol6}) on the current difficulties of reproducing deployments with custom code~\cite{boettiger2015introduction}. In fact, the idea of deploying a custom lightweight container to the client, which contains a custom sandboxed environment, seems promising. While this would arguably make custom deployments take more time, it would still be incomparably faster than the standard workflow for updating the client installation, as that is a decision the user cannot take by themselves and may take days or weeks, depending on organizational processes.

\section{Conclusion}
\label{conclusion}
OODIDA was originally designed with the goal of enabling rapid prototyping, largely achieved by automated deployment of the client installation. In reality, however, organizational bureaucracy and the demands of industry best-practices slow down this process a lot. In contrast, with the help of the active-code replacement feature, we provide a safe and easy-to-use alternative that sidesteps such hurdles. Of course, we had to make some concessions by placing limitations on user-defined custom-code modules. Nonetheless, active-code replacement is eminently useful in practice as it enables data analysts working with OODIDA to pursue a highly interactive workflow, given how quickly custom code can be deployed on the system.

\subsubsection*{Acknowledgements.} 
This research was financially supported by the project Onboard/Offboard Distributed Data Analytics (OODIDA) in the funding program FFI: Strategic Vehicle Research and Innovation (DNR 2016-04260), which is administered by VINNOVA, the Swedish Government Agency for Innovation Systems. It took place in the Fraunhofer Cluster of Excellence "Cognitive Internet Technologies." Ramin Yahyapour (University of G{\"o}ttingen) provided insightful comments during a poster presentation.

%
%
%
%

\end{document}